\documentclass[aps,prl,twocolumn,showpacs,floatfix]{revtex4}
\usepackage{graphicx,color}
\usepackage{amsmath}

\newcommand{\dbar}{{\bar{d}}}

\newcommand{\figone}{\begin{figure}[htbp]
    {\includegraphics[width=1.67in,clip]{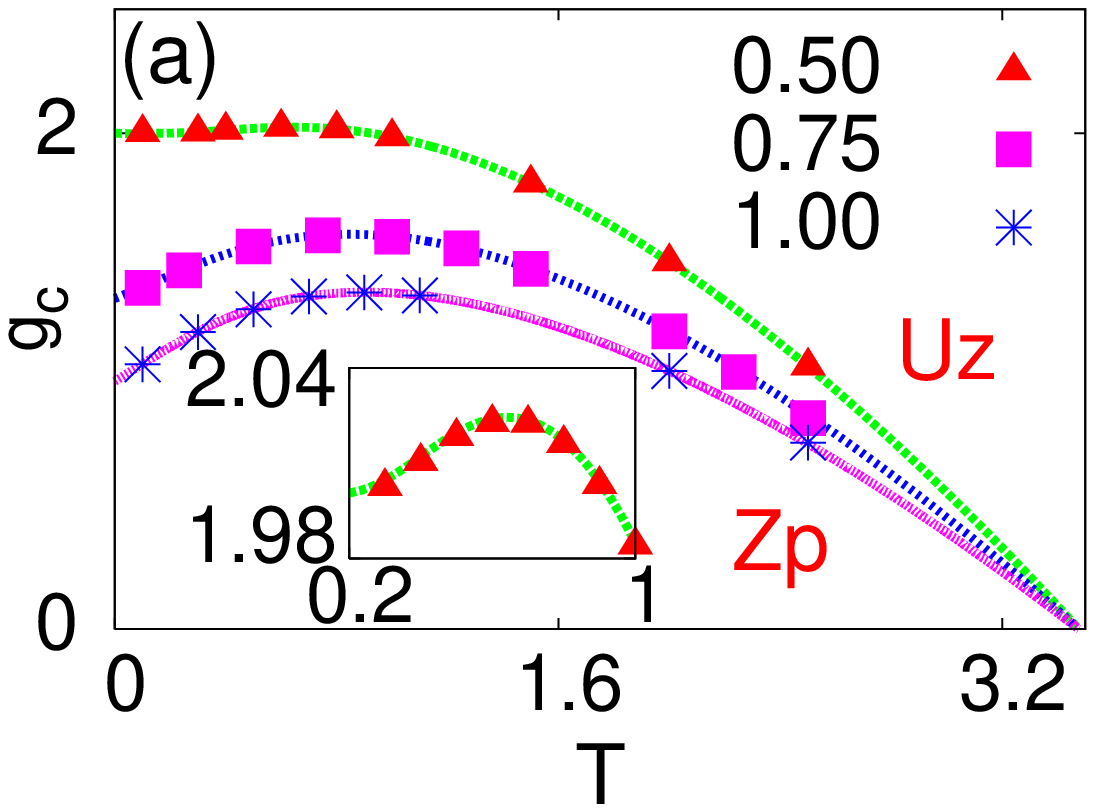}}
   {\includegraphics[width=1.67in,clip]{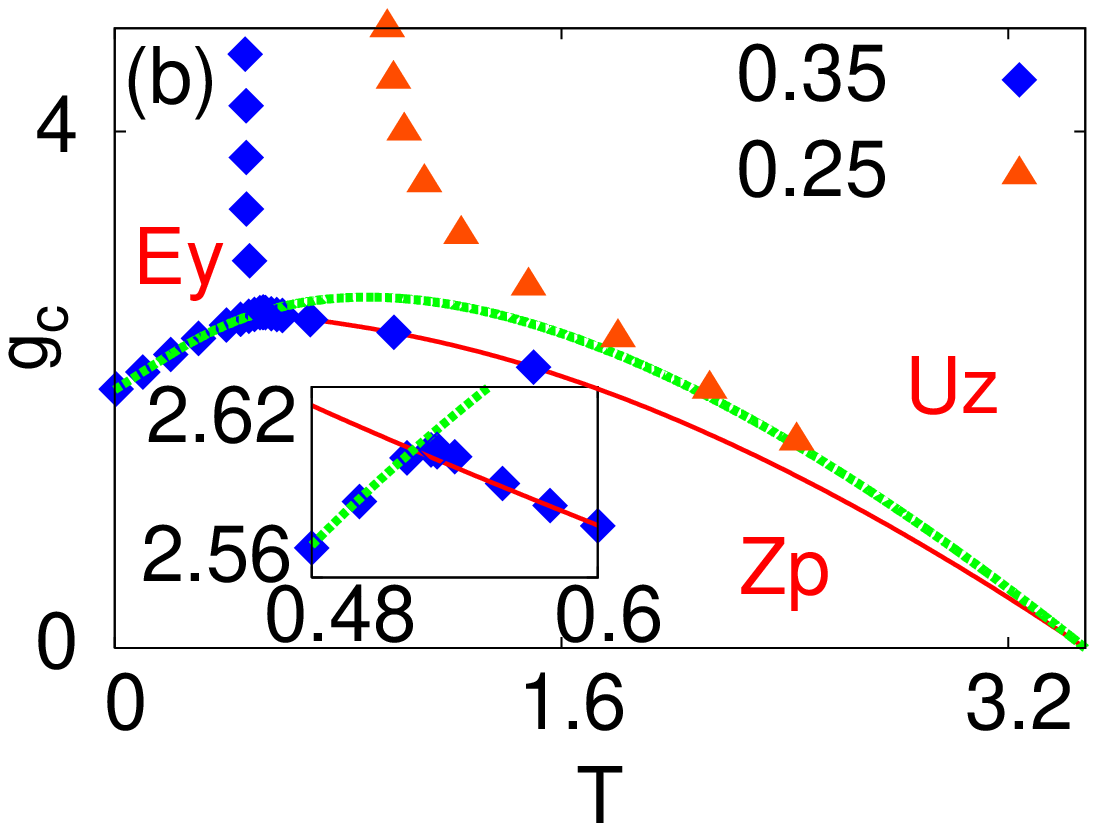}}
\caption{ The $g_{c}$ versus $T$ phase diagram
  in the fixed force ensemble. Lines are the exact results while the
  points are from numerics.  (a) For $0.5 \le s \le 1$.  The inset
  shows the re-entrant region for $s = 0.5$.  (b) For $s=0.35$ and
  $s=0.25$. The inset shows the triple point for $s=0.35$.}
\label{fig:phase1}
\end{figure}
}

\newcommand{\figsch}{\begin{figure}[htbp]
    {\includegraphics[height=1.in,clip]{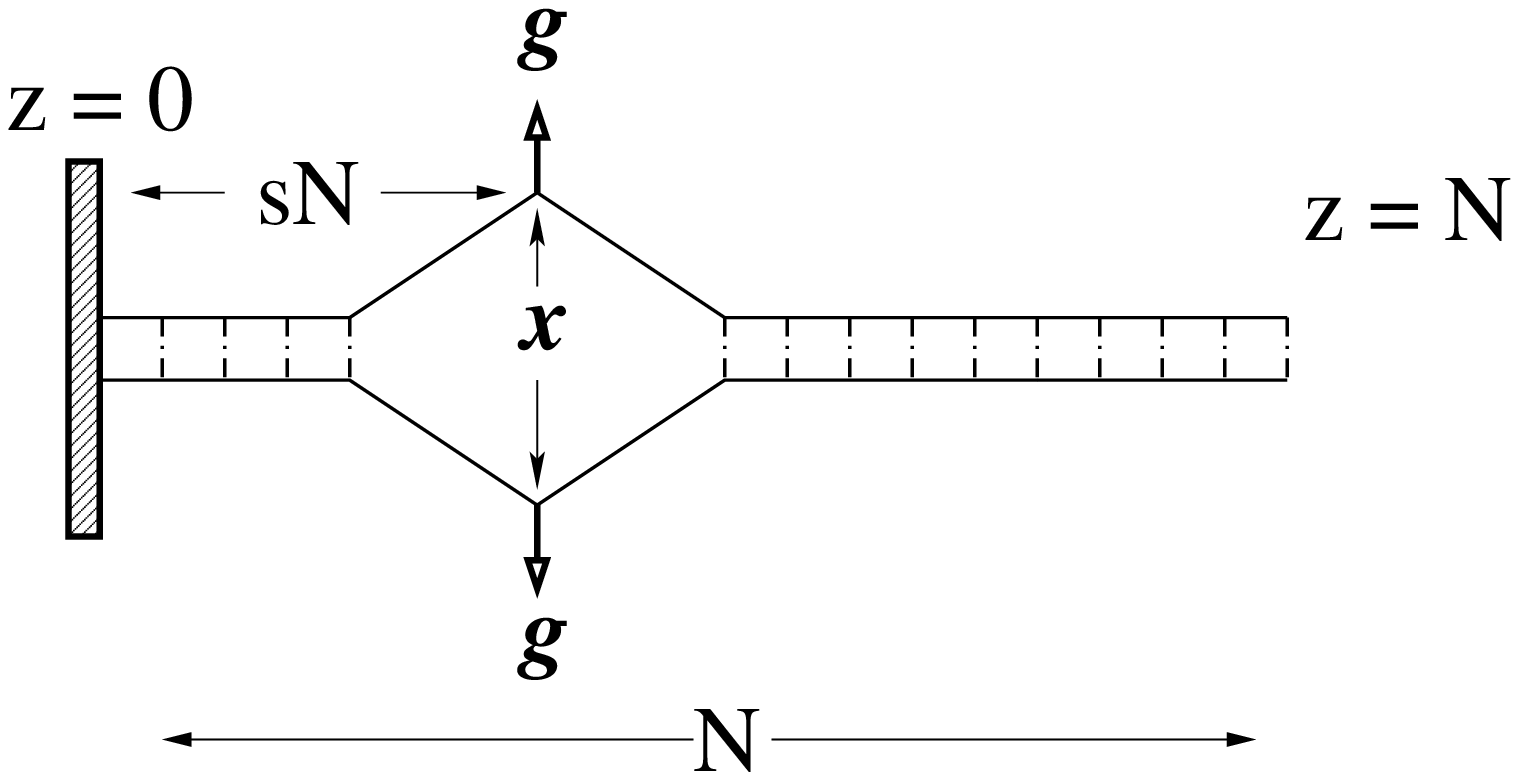}}
\caption{ Schematic diagram of DNA unzipping by a pulling force at a fraction
  $s$ ($0 \le s \le 1$) from the anchored end. In the fixed force
  ensemble the force {$g$} is kept fixed while the separation {$x$} is
  kept fixed in the fixed distance ensemble. }
\label{fig:schdia}
\end{figure}
}

\newcommand{\figeyeone}{\begin{figure}[t]
   {\includegraphics[width=1.67in,clip]{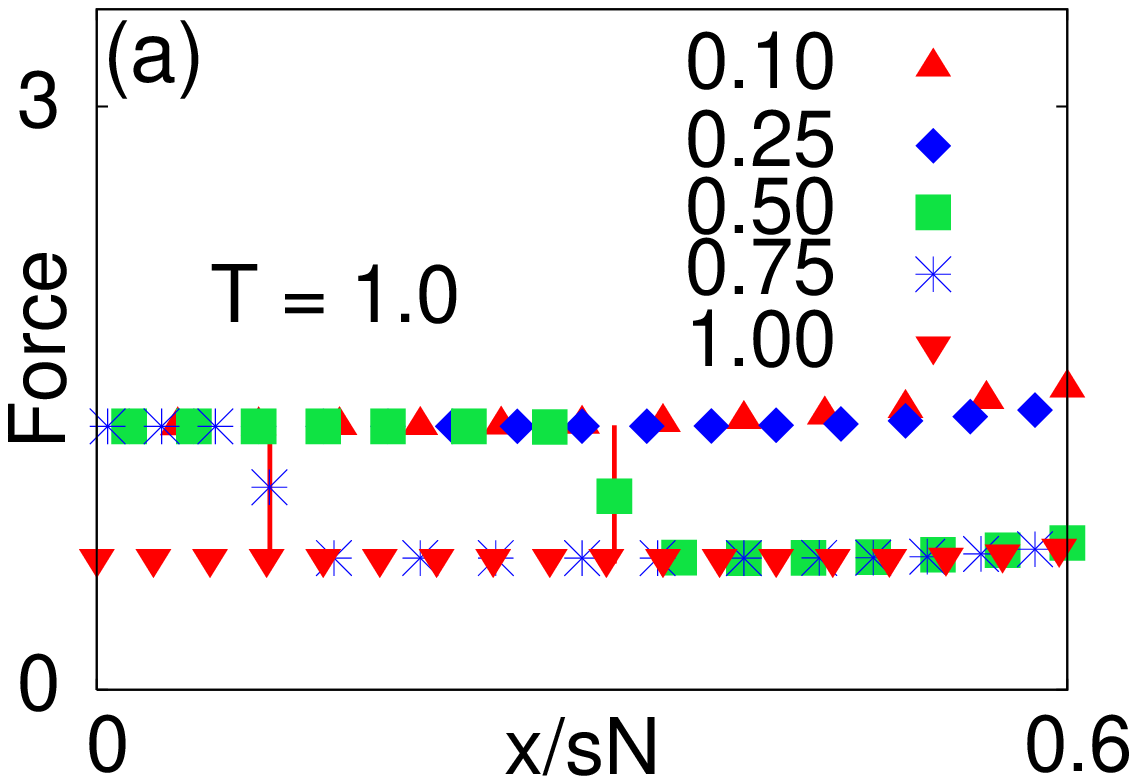}}
   {\includegraphics[width=1.67in,clip]{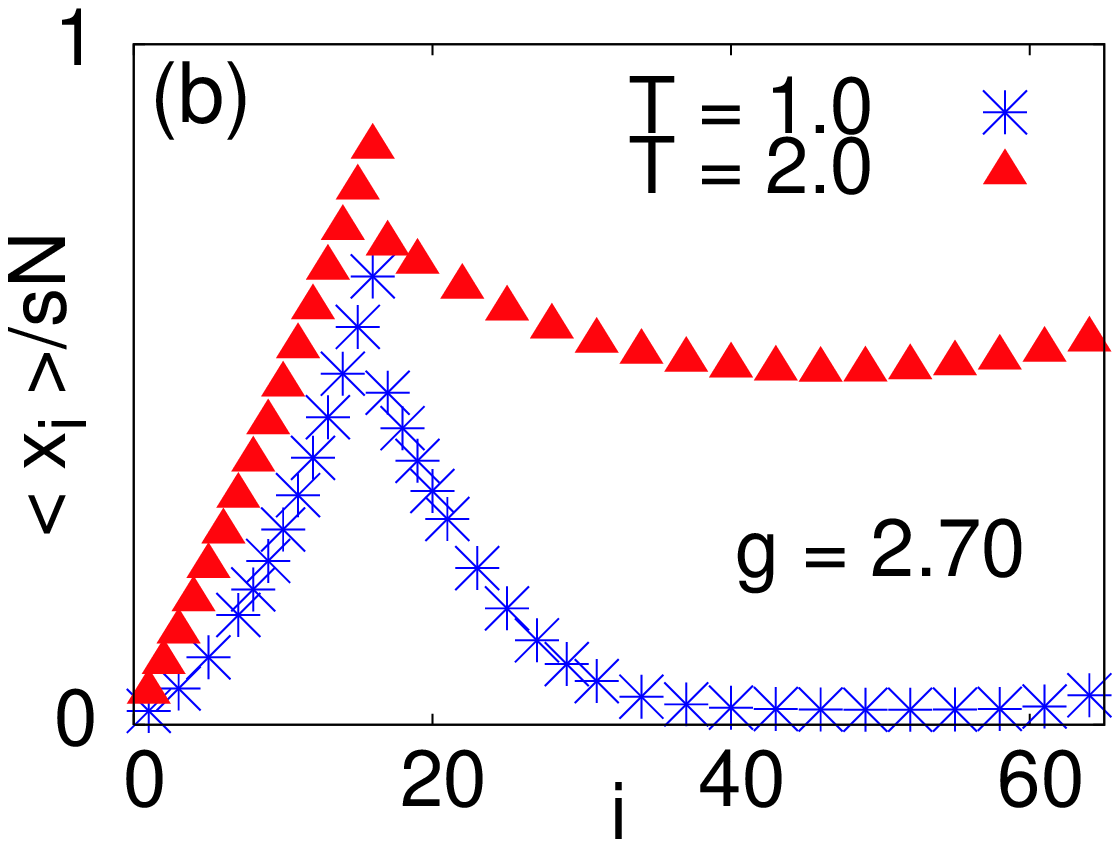}}
\caption{ (a) Scaled separation $x/(sN)$ versus force isotherm for
  different $s$ at $T = 1.0$ with $N=256$ in the fixed distance
  ensemble.  The location of $X(s,T)$ (see text) is shown by the solid
  line.  
 (b) Average separation (fixed force ensemble) between the
  ${i}^{th}$ monomer of the DNA of length $N=64$ for T = 1.0 for a force
  $g = 2.7$ at $s=0.25$ at two different temperatures. }
\label{fig:eye}
\end{figure}
}

\newcommand{\figdiey}{\begin{figure}[t]
   {\includegraphics[width=3.0in,clip]{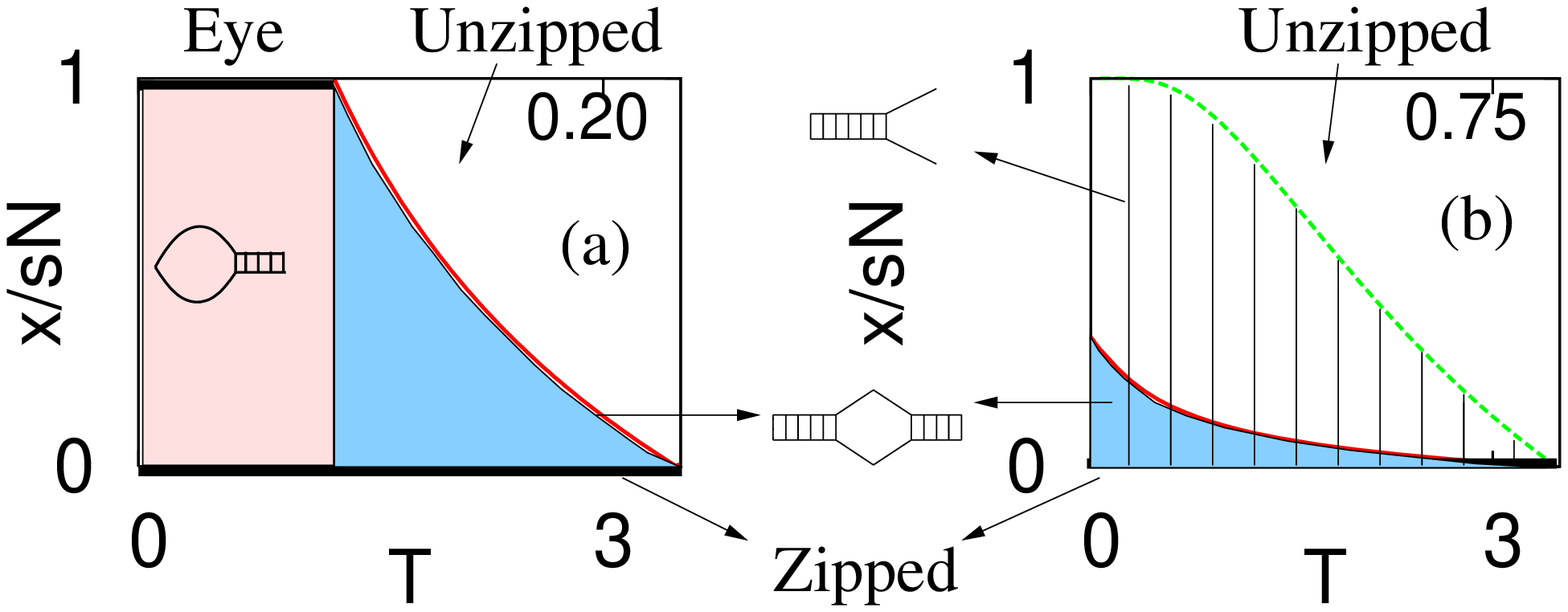}}
\caption{  Fixed distance ensemble  $T$ vs ${x}/{(sN)}$ phase diagram
  for (a) $s=0.2$ and (b) $s=0.75$ .  The zipped and the eye phases
  are shown by thick lines.  The coexistence regions are marked by
  different shades or vertical lines.   $X_c(T)$, defined after
  Eq. (\ref{eq:4}),  is represented by the dotted line in (b). 
}
\label{fig:fixdis1}
\end{figure}
}

\begin{document}
\bibliographystyle{prsty}      

\title{Complete Phase Diagram of DNA Unzipping: Eye, Y-fork and
  triple point} 
\author{ Rajeev  Kapri$^{1*}$, Somendra M.~Bhattacharjee$^{1*}$ and Flavio Seno$^{2}$}
\email{rajeev@iopb.res.in,         somen@iopb.res.in,   flavio.seno@pd.infn.it }
\affiliation{$^1$ Institute of Physics, Bhubaneswar 751 005, India\\
  $^2$ INFM, Dipartimento di Fisica, Universit\'a di Padova, Via
  Marzolo 8,35131 Padova, Italy}

\begin{abstract}
  We study the unzipping of double stranded DNA (dsDNA) by applying a
  pulling force at a fraction $s$ $(0 \le s \le 1)$ from the anchored
  end.  From exact analytical and numerical results, the complete
  phase diagram is presented.  The phase diagram shows a strong
  ensemble dependence for various values of $s$.  In addition, we show
  the existence of an ``eye'' phase and a triple point.
\end{abstract}
\pacs{87.14.Gg, 82.35.Pq,64.70.-p,36.20.Ey }
\date{\today}
\maketitle

The initial step in DNA replication and RNA transcription, as the
enzyme associates with DNA, is to open a few base pairs near it. In
case of replication this opening takes place near one of the ends,
whereas for transcription it can be anywhere on the
DNA\cite{alberts03,cheetham99,cook}.  The ubiquity of the process
calls for a mechanism that does not require high temperatures or
extreme pH conditions, unlike melting to which it generally gets
associated.  One such possibility is a force induced unzipping
transition\cite{somen:unzip}.  This transition has now been well
established both theoretically
\cite{somen:unzip,sebastian,maren:phase,monte,scaling,zhou,chen,lubensky:prl,kafri,lam}
and experimentally\cite{danil,ansel,essevaz:micro}.  The
focus of attention so far has been the geometry reminiscent of the DNA
replication, pulling only the open end of dsDNA. However,
transcription requires pulling DNA at an intermediate point, often
with DNA getting anchored to cytoplasmic membrane\cite{cook} {\it in
  vivo} .   Similarly,  end constraint is  important  in
circular DNA as, e.g., in bacteria like E. Coli. 
Anchoring of one end is also used in single molecule
experiments\cite{danil,ansel}.  The richer surprises in this type of
geometry provide the primary motivation for working out the full
unzipping phase diagram.

The extensive theoretical
work\cite{somen:unzip,sebastian,maren:phase,monte,scaling,zhou,chen,lubensky:prl,kafri,lam}
on the unzipping transition in various avatars of the basic Poland and
Scheraga model \cite{poland} and the nature of the real phase
diagram\cite{danil} recently obtained for lambda phage DNA indicate
that the basic features are preserved in simpler exactly solvable
models\cite{maren:phase,scaling} {\it even in two dimensions}.  These
basic results include the first order nature of the unzipping
transition and the existence of a re-entrant region allowing unzipping
by decreasing temperature.  In this paper our aim is to study the
force induced transition on the lattice model used previously in Ref.
\cite{scaling} but with the pulling force applied on a base pair which
is $N_1$ monomers away from the ancorhed end of the DNA molecule (of
total length $N$).  We call $s$ the fixed fraction $N_1/N$.  See Fig.
\ref{fig:schdia}.  In single molecule experiments, the results may
depend on the statistical ensemble used\cite{maren2,busta}.  One may
recall that instruments like atomic force microscopes\cite{ansel} use
the fixed distance ensemble while the magnetic bead method of Ref
\cite{danil} uses the fixed force ensemble.  Therefore, we studied the
unzipping transition both in the fixed force and the fixed distance
ensembles, by using analytical and exact transfer matrix methods,
though we concentrate mostly on the fixed force case in this paper.

\figsch 

Before describing the model, let us point out   a few of the basic
results we have obtained.  The phase diagrams in both the fixed distance
and fixed force ensembles are obtained.  The qualitative features of
the phase diagram, especially the nature of the phases, the
s-dependence and the ensemble dependence are generic as can be shown
by general arguments.  An ``eye''-like configuration exists for all
$s<1$ in the fixed distance ensemble, either as a distinct phase (when
it extends upto the anchored end) or as two ``Y'''s joined together
(see fig.  \ref{fig:fixdis1}) where a ``Y'' is a coexistence of the
unzipped and the zipped phases.  These configurations resemble the
``Y'' fork in replication and the transcription bubble.  In a fixed
force ensemble, an eye exists only for low values of $s$ ($s<1/2$ in
our model) and the phase diagram depends on $s$.  For values of $s$
where the ``eye'' phase exists, there is triple point at the
intersection of the zipped-eye (Zp-Ey), eye-unzipped (Ey-Uz), and
zipped-unzipped (Zp-Uz) phase boundaries.

The model is defined as follows.  The two strands of a homo-polymer DNA are
represented by two directed random walks on a $d=2$ dimensional square
lattice. The walks, starting from the origin, are restricted to go
towards the positive direction of the diagonal axis (z-direction) {\it
  without crossing each other}. There is a gain of energy $-\epsilon$
($\epsilon >0$) for every contact (i.e. separation $x=0$). The
directional nature of the walks takes care of the correct base pairing
of DNA.  In addition to this bonding, a force $g$ acts along the
transverse direction ($x$-direction) at a fixed fraction $s$ ($0 \leq
s \leq 1$) from the anchored end ($z=0$). As is well-known, this force,
though acts at a point, affects the bulk behavior\cite{somen:unzip}. 
The quantities of interest
depend on the ensemble one is working with.  For example, it is the
average separation $\langle x \rangle $ at the point of application of the
force in the fixed force ensemble, whereas, in the fixed distance
ensemble, it is the average force $\langle g \rangle$ needed to
maintain the distance $x$ between the two strands. $(g,x)$ constitute
a thermodynamic conjugate pair.

A taste of the surprise for $s\neq 1$ can be gleaned from a simple
analysis that is exact in the low temperature region.  For $s=1$, this
argument gives the exact reentrant phase
boundary\cite{scaling,maren:phase,monte}.  If $s<1/2$, at $T=0$, the
force opens the chain maximally so as to form an eye (extensive in
length).  The energy $E$ and entropy $S$ of the eye with respect to
the completely zipped chain are $E=-gasN + 2 \epsilon sN$ and $S=- 2sN
\ln \mu_b$, where $\mu_b$ is the connectivity constant of the bound
phase and $a$ is a geometric factor.  Throughout this paper we take
the Boltzmann constant $k_{B}=1$.  For our lattice model, $\mu_b=2$
and, we set $a=1$  by choosing  the elementary diagonal of the
underlying square lattice as the unit of length.   A transition from the 
zipped (Zp)
state is therefore possible if $g>g_c(s,T)=2 (\epsilon + T \ln
\mu_b)/a$ which is double the force required for the unzipping
transition at $s=1$.  The situation is different for $s\geq 1/2$,
where complete unzipping is possible at $T=0$.  In the completely
unzipped (Uz) state, the energy and entropy with respect to the zipped
state are $E= N \epsilon -gasN$ and $S=- N\ln \mu_b+ 2 (1-s)N \ln
\mu_1$ where $\mu_1$ is the connectivity constant for a single chain.
The low temperature phase boundary is given by $g_c(s,T)= (\epsilon +T
\ln \mu_b/\mu^{2(1-s)}_1)/(sa)$.  For our lattice problem $\mu_1=2$.
Hence, there will be no low temperature re-entrance if $s=1/2$.   These
$s$-dependences match the exact results.

To trace out the exact phase boundary we use the recursion relation
method.  Let $D_t(x,x^{\prime})$ be the partition function with
separations $x$ and $x^{\prime}$ at the two extreme ends of a dsDNA of
length $t$. Then,
\begin{eqnarray}
  \label{eq:1}
D_{t+1}(x,x^{\prime})&=&[2 D_t(x,x^{\prime}) + D_t(x,x^{\prime}+1)\nonumber\\
&&\quad\qquad +D_t(x,x^{\prime}-1)](1+P\delta_{x^{\prime},0}), 
\end{eqnarray}
with $P=e^{\beta}-1$, $\beta=1/T$, and initial conditions
$D_0(x,x^{\prime})=\delta_{x,x^{\prime}}$.  Mutual exclusion is
ensured by $D(x,x^{\prime})=0$ whenever any (or both) of the two
arguments $x,x^{\prime} < 0$.  One can construct two other partition
functions, (i) $d_t(x)\equiv D_t(0,x)$ when one end is held fixed, and
(ii) $\dbar_t(x)\equiv \sum_{x^{\prime}}D_t(x,x^{\prime})$ when one
end is free.  Of these, $d_t(x)$ has been used in previous studies of
the force at the end case\cite{scaling} where the phases and the
transitions come from the singularities of the generating function,
${\cal G}(z,\beta,g) = \sum_{t} \sum_x z^t e^{\beta gx} d_t(x)$.
These singularities are $z_1=1/4$, $z_2=(2+2\cosh{\beta g})^{-1}$ and
$z_3=\sqrt{-e^{-\beta} +1} - 1 + e^{-\beta}$.  The zero force melting,
coming from $z_1=z_3$, is at $T_c=1/\ln(4/3)=3.476059497...$.  The
unzipping phase boundary 
can be determined in the fixed distance ensemble by noting that the
force required to maintain the 
separation $x$ is $-T\partial \ln d_N(x) /\partial x$.  By using
$d_N(x) \approx \lambda(z_3)^x/z_3^{N+1}$ for large $N$ with
$\lambda(z)=(1-2z-\sqrt{1-4z})/(2z)$, one gets
\begin{equation}
  \label{eq:4}
g_c(T)\equiv g_c(s=1,T) = - T \ln \lambda(z_3),\quad {\mathrm (Zp
    \Leftrightarrow Uz)}.
\end{equation}
This  is the known $s=1$ phase boundary\cite{scaling} coming from
$z_2=z_3$ in the fixed force ensemble.  On this boundary, the end
separation is given by $X_c(T)\equiv x/N= \tanh [g_c(T)/T]$.  The
phase coexistence on this boundary gives the Y-fork structure, which
we simply call a ``Y''.

\figdiey

{\bf Fixed distance ensemble:}
 If the distance or separation of the two
strands at $t=sN$ is kept fixed at $x$, while the DNA is anchored
($x=0$) at $t=0$ but free at the other end at $t=N$, the partition
function is  ($z_3$ dependence of $\lambda$ suppressed)
\begin{eqnarray}
\label{eq:2}
Z_N(x,s)&=& d_{sN}(x) \overline{d}_{(1-s)N}(x)\nonumber\\
&\approx&
{\lambda^x}{z_3^{-sN}} \left(
4^{(1-s)N}+{\lambda^x}{z_3^{-(1-s)N}}\right ).
\end{eqnarray}
In the limit
$N\rightarrow\infty$ for a fixed $s$, the larger of the two terms (in
the big parenthesis on the right hand side) contributes to the free
energy.  For 
\begin{equation}
  \label{eq:6}
  \frac{x}{sN} < X(s,T)\equiv \frac{1-s}{s} \frac{\ln (4z_3)}{\ln (\lambda(z_3))},
\end{equation}
the larger term is the second one, otherwise it is the first one.
Therefore, we get a phase boundary
\begin{subequations}
\begin{eqnarray}
  \label{eq:5}
  g_c(s,T)&=&2g_c(T), \ {\rm if\ Eq. \ref{eq:6}} 
  \qquad   ({\rm Zp}  \Leftrightarrow {\rm Ey})\\
\label{eq:5b}
 &=& g_c(T), \ {\rm otherwise},
      \quad ({\rm Zp} \Leftrightarrow {\rm Uz}).
\end{eqnarray}
\end{subequations}
With the increase of the separation $x$, the end point gets detached
at the critical value $x=sNX(s,T)$, provided $X(s,T)<1$.  Once all the
bonds are broken the two open tails behave like free independent
chains.  In such a situation, the force required to maintain the
separation is just like the $s=1$ end case (in the
$sN\rightarrow\infty$ limit) as we see in Eq. \ref{eq:5}.  For this to
happen we also require $X(s,T)<X_c(T)$, else the DNA will be in the
unzipped phase.  Fig.~\ref{fig:fixdis1} shows the phase diagram on a
temperature-distance plane for two values of $s$ in the fixed distance
ensemble, though in the force-temperature plane the phase boundaries
are independent of $s$ as follows from Eqs. \ref{eq:5},\ref{eq:5b}.
An ``eye'' of the type shown in Fig ~\ref{fig:schdia} occurs in the
coexistence region shown by solid vertical lines in Fig.
\ref{fig:fixdis1} and is to be interpreted as two ``Y'''s joined
together.  This configuration is analogous to the transcription bubble
produced e.g., by RNA polymerase\cite{cheetham99} a subunit of which
keeps the two strands of DNA separated.

To supplement the exact results, the partition function for the two
strands starting from origin is obtained numerically by using the
exact transfer matrix technique for the recursion relation Eq.
(\ref{eq:1}).  In the fixed distance ensemble, the distance between
the two strands $x$ is varied at a step of $1$ (length of diagonal of
the square lattice) from $0$ (``zipped'') to $sN$ (``completely
stretched''). The quantity of interest, the average force required to
maintain the distance $x$ between two chains, is calculated by using
finite differences in free energy.  The scaled separation between the
two strands of DNA, $x/sN$, versus the corresponding average force at
$T=1.0$ is shown in fig. \ref{fig:eye}a, for several values of
$s$. Fig \ref{fig:eye}b shows the eye formation in the fixed force
ensemble and is discussed next.
When the end monomers of the chains ($s=1.0$) are maintained at a
fixed distance, there is only one plateau at the critical force given
by Eq. (\ref{eq:4}) (the Zp-Uz phase boundary).  When $s<1.0$, the
force-distance isotherm has two plateaus as per Eqs. (\ref{eq:5}),
(\ref{eq:5b}) with a step that matches with the critical value
$X(s,T)$.  We do not go into further details of this phase diagram
here because the subtleties are more prominent in the fixed force
ensemble.  

\figeyeone

{\bf Fixed force ensemble:} In the fixed force ensemble, the
generalization of the generating function defined for the end case
below Eq. (\ref{eq:1}) is ${\cal G}_s(z,\beta,g)$ = $ \sum_{x>0}
e^{\beta g x} \sum_t z^t Z_t(x,s)$.  Using Eq. (\ref{eq:2}), this can
be written as
\begin{equation}
\label{eq:13}
{\cal G}_s(z,\beta,g)\approx \sum_x e^{\beta g x} \{[\lambda(4^{(1-s)/s} z^{1/s})]^x + 
        [\lambda(z)]^{2x} \},
\end{equation}
where the first term on the right hand side in curly bracket
represents the unzipped state while the second term is for the eye
state.  The additional singularities in $z$ are then
$z_2=[{4^{1-s}[2(1+\cosh \beta g)]^s}]^{-1}$, which goes over to the
$z_2$ mentioned earlier for $s=1$, and $z_4=\{2[1+\cosh (\beta
g/2)]\}^{-1}$.  The zipped to unzipped phase transition occurs at
$z_2=z_3$ while a zipped to eye phase transition takes place at
$z_3=z_4$.

\figone

The $s$ dependence of the singularities show that there cannot be an
eye phase in the fixed force ensemble if $s\geq 1/2$, even though one
may open an eye of the type of Fig. \ref{fig:schdia}  in the fixed
distance ensemble. 
In this situation, only transition possible is unzipping with an
$s$-dependent boundary given by
\begin{equation}
\label{eq:3}
g_c(s,T)= T \ln \lambda(4^{(1-s)/s} z_3^{1/s}) \quad ({\rm Zp}
    \Leftrightarrow {\rm Uz})
\end{equation}
which matches with $g_c(T)$ for the end case.  In addition, close to
$T=0$, we see $g_c(s,T)\approx \frac{\epsilon}{s} +\frac{2s-1}{s} \ln
2$ which corroborates the results from the simple argument, including
the vanishing slope at $T=0$ for $s=1/2$ ( the absence of a
low temperature re-entrance).  The inset in Fig. \ref{fig:phase1}a
shows that there is still a small region in the intermediate
temperature range where one does see a re-entrance.

With a force at a point $s<1/2$, a phase boundary comes from $z_3=z_4$
which matches with the boundary we already derived in Eq.
(\ref{eq:5}).  This is the transition to eye in presence of a force,
as found in the fixed distance ensemble also.  However the eye phase
cannot continue for the whole temperature range, definitely not at the
melting point where the unzipped phase should be recovered.  The
unzipped phase boundary comes from $z_2=z_4$ which yields Eq.
(\ref{eq:3}).  The lower of the two curves would determine the
thermodynamic phase boundary.  See Fig.  \ref{fig:phase1}.  These
results also suggest the possibility of an eye to unzipped phase
transition which however eludes the approximation done in Eq.
(\ref{eq:13}) (based on Eq.  (\ref{eq:2})). This phase boundary is
determined numerically below.  The intersection of the three
boundaries, which occurs only for $s<1/2$, is a {\it triple point}.

To extend the analysis for $s<1/2$, and to verify the analytical
results, we use the exact transfer matrix technique, mentioned
already, but now with an applied force.  The average separation
$\langle x\rangle$ between the two chains at the site of application
of force at a temperature $T$, is calculated by taking the finite
difference of exact free energies as $g$ is increased in steps of
$\Delta g = 0.001$.  The critical force in the thermodynamic limit is
determined by $N\rightarrow\infty$ extrapolation of the crossing
points of $\langle x\rangle$ versus $g$ curves for pairs of length
$N$. We take $N$ from $400$ to $1000$.  The results are shown in Fig.
\ref{fig:phase1}, with very good agreement with the analytical
results.

The maximum separation between the chains at the point of application
of the force  is $sN$, and at this separation, the end monomers are always in an
unzipped state  for $0.5 \le s \le 1$ even for $T\rightarrow 0$.  This
is not so for $s<0.5$.  In this case, the end monomers will of course
be unzipped at higher temperatures but that does not rule out the
possibility of a zipped phase at low temperatures.
To study how the end monomer separation behaves with $T$
when $s < 0.5$, the force is fixed to a value which lies on the phase
boundary obtained above and $T$  is increased. At low
temperatures, we find that the end monomers are in zipped state (i.e.
in contact) even when the separation between the two chains, where
force is applied, is maximum. At a particular temperature, which
depends on the fraction $s$, the end monomer separation becomes
macroscopic, signaling an Ey-Uz phase transition.  
Fig.  \ref{fig:phase1} shows the phase boundary obtained by repeated
application  of this procedure..

The complete phase diagram for $s=0.35$ and $0.25$ are shown in Fig.
\ref{fig:phase1}b. The three  coexistence lines meet at a
triple point $T_{p}$ which shifts with $s$. The triple point moves
towards the high temperature side when $s$ is decreased and merges
with the melting point for $s=0.25$. The inset shows the details
around the triple point.
The limit $s\rightarrow 0$ is singular because the eye phase cannot exist and
a force applied at the anchored point cannot open a chain.  In
that limit, the Ey-Uz boundary becomes vertical (parallel to the force
axis) and is the only meaningful phase boundary.

The Ey-Uz transition can be seen in Fig. \ref{fig:eye}b which shows
the average separation between the monomers of the DNA of length
$N=64$ with a force at $s=0.25$.  We calculate the separation for at
two different temperatures, both at $g=2.7$ which lies on the phase
boundary of the zipped and the eye phases at $T=1$, and another for
the same force but at $T=2$ deep in unzipped region (see
fig.~\ref{fig:phase1}).  The monomers in the outer most part of the
DNA are in the zipped state for $T=1$ and $g=2.7$ showing the
formation of the eye but for $T=2$ the DNA is in the unzipped phase.
In the limit of $N\rightarrow\infty$ the number of bound pairs at the
end of the chain is extensive ($\propto N$) in the eye phase but not
in the unzipped phase.

We may now summarize some of the results of this paper.  (1) The phase
diagrams in the fixed distance and fixed force ensemble are shown in Figs.
\ref{fig:fixdis1} and \ref{fig:phase1} which also show when the
``eye'' appears as a distinct phase  or as two ``Y'''s.
(2) There is a low temperature
re-entrance in the fixed force ensemble for all values of $s$ except
for $s=1/2$.  In the latter  case the re-entrance is restricted to a small
intermediate temperature range.  (4) In the fixed
force ensemble, the phase boundary shifts with $s$ as it is decreased
from $s=1$ to $s=1/2$.   In this range there is no triple point.  (5)
For $s<1/2$, the low temperature phase 
boundary representing the zipped-to-eye phase transition in the fixed
force case is independent of $s$ and it intersects the $s$-dependent
zipped-to-unzipped boundary at a triple point. There is an additional
eye to unzipped phase transition line in the large force regime.  (6)
The triple point shifts towards the zero-force melting point $T_c$ as
$s$ is decreased. 
Although our approach is based on a coarse grained model, we believe
that our results are robust to be observed by high precision
measurements of DNA unzipping under a pulling force.
 
We  thank Sanjay Kumar for suggesting the relevance of $s$ in
unzipping and acknowledge the support from PRIN-MURST 2003
and FISR 2001.

\end{document}